\documentclass[sigconf, screen]{acmart}
\renewcommand\footnotetextcopyrightpermission[1]{} 

\usepackage{fancyhdr}
\usepackage{pifont}
\usepackage{amsmath}
\usepackage[dvipsnames]{xcolor}
\usepackage{tcolorbox}
\usepackage{makecell}
\usepackage{afterpage}
\usepackage{hyperref}

\definecolor{mygreen}{RGB}{0,147,0}
\definecolor{myred}{RGB}{150,0,0}
\usepackage{enumitem}

\setcopyright{none}

\author{Yi Qian$^{1}$, Kunwei Qian$^{1}$, Xingbang He$^{1}$, Ligeng Chen$^{2}$, Jikang Zhang$^{3}$,Tiantai Zhang$^{1}$, Haiyang Wei$^{1}$, Linzhang Wang$^{1}$, Hao Wu$^{1,*}$, Bing Mao$^{1}$}
\affiliation{
  \institution{$^{1}$State Key Laboratory for Novel Software Technology, Nanjing University}
  \institution{$^{2}$Hornor Device Co., Ltd}
  \institution{$^{3}$Institute of Dataspace, Hefei Comprehensive National Science Center}
  \country{China}
}
\email{{yi\_qian, kunweiqian,xingbanghe,chenlg,jikangzhang,weihaiyang}@smail.nju.edu.cn}
\email{z472421519@gmail.com, {lzwang,hao.wu, maobing}@nju.edu.cn}
\thanks{*Corresponding author.\\This paper has been accepted to appear in the Proceedings of the 33rd ACM Conference on Computer and Communications Security (CCS '26).}

\begin{document}

\title{Mind the Gap: Action Rebinding Attacks \\against Android GUI Agents}

\begin{abstract}
Large multimodal model powered GUI agents are emerging as high-privilege operators on mobile platforms, entrusted to perceive screen content and inject inputs across application boundaries. While these agents aim to automate complex tasks, we demonstrate that their design introduces a fundamental conflict with Android's strict application sandboxing. We present a novel cross-application \textbf{Action Rebinding attack}, which allows a malicious application with zero dangerous permissions to hijack the agent's execution and perform privileged operations on behalf of the attacker.

Our attack exploits the inevitable observation-action gap inherent in the agent's reasoning pipeline. A malicious app can render a benign ``contextual carrier'' to elicit a planned action, and then swap the foreground to a sensitive target application during the reasoning latency. The agent, unaware of the transition, unwittingly executes the action in the privileged context. We further advance this attack by weaponizing the agent's own task-recovery logic to create programmable, multi-step exploit loops , and introducing an Intent Alignment Strategy (IAS) that manipulates the agent's reasoning to rationalize the hijacked state.

We evaluate our attack on six widely-used Android GUI agents. Our results demonstrate a 100\% success rate for atomic action hijacking and the ability to orchestrate high-impact exploits, including unauthorized file deletion, SMS transmission, and app uninstallation, without the attacker holding any corresponding permissions.  Furthermore, since the malicious application separates intent from capability and contains no privileged API calls, it achieves a 0\% detection rate across commercial malware scanners (e.g., VirusTotal), highlighting a critical blind spot in current mobile security analysis. To access experimental logs and demonstration videos, please contact \textit{yi\_qian@smail.nju.edu.cn}.
\end{abstract}



\maketitle

\section{Introduction}

The integration of Large Multimodal Models (LMMs) into mobile operating systems has catalyzed a paradigm shift in human-computer interaction: the transition from user-driven application usage to autonomous \textit{GUI agents}. 
Unlike traditional automation restricted to specific APIs, these agents function as general-purpose assistants that interact with the system through the shared user interface layer~\cite{zhang2025appagent,jiang2025appagentx,ye2025mobile,liu2024autoglm,zhang2025mobiagent,mobileuse,droidrun}.
They perceive the screen visually, decompose high-level user commands into a sequence of discrete UI interactions, and inject system-wide input events to execute complex, cross-application tasks~\cite{zhang2024llamatouch, liu2024autoglm}. 
In this emerging ecosystem, GUI agents effectively serve as a new \textit{control plane} that resides between the user and the underlying operating system, mediating interactions across otherwise isolated application boundaries.

\begin{figure}[h]
  \centering
  \includegraphics[width=\linewidth]{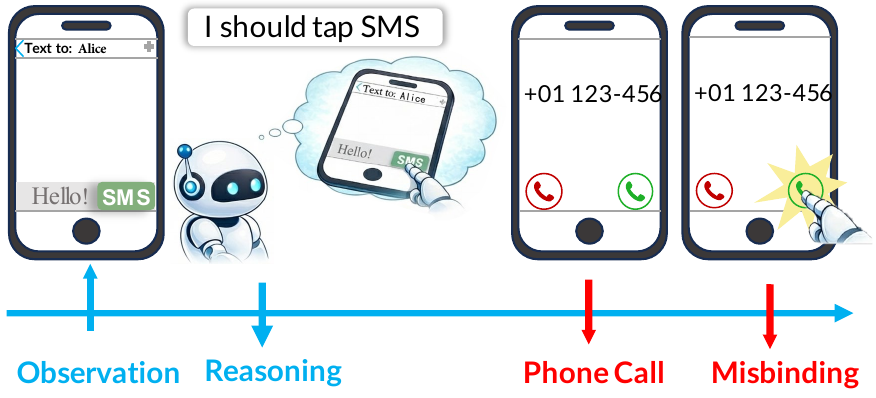}
  \caption{The agent's action to send a message is misbound to answering the phone  due to an unexpected phone call.}
  \Description{The agent's action to send a message is misbound to answering the phone  due to an unexpected phone call.}
  \label{fig:intro}
\end{figure}

These agents follow an \textit{observation-reasoning-action} pipeline. 
They perceive visual information, infer the next action, and execute on the UI layer.
This pipeline operates under the implicit assumption of \textbf{Visual Atomicity}: \textit{the UI state captured during the observation stage remains invariant until the subsequent action is executed.}
However, we observe that such atomicity is not guaranteed in the event-driven Android environments.
Due to either intentional application behavior (e.g., proactive foreground transitions) or asynchronous system events (e.g., incoming calls or notifications), the UI state can undergo unpredictable mutations within the \textbf{temporal window} between an agent's perception and its action injection. 
Consequently, the agent's action may be misapplied to unintended UI components, leading to \textit{action misbinding} (Figure~\ref{fig:intro}).

Exploiting this inherent \textit{observation-to-action} gap, we introduce a novel attack surface: \textbf{Action Rebinding}.
By manipulating the UI state during this gap, an attacker can redirect the agent's intended action toward a target component of the attacker's choice.
We identify two architectural properties of the Android OS that significantly amplify this threat:\textbf{(1) Unconstrained Foreground Transitions:} Android allows any application to initiate a foreground transition via \texttt{Intent}s without requiring privileged permissions. Attackers can weaponize this to proactively alter the UI context.
\textbf{(2) UI State Preservation:} The OS preserves the internal state of applications during background-foreground transitions. This allows attackers to maintain a persistent exploit context across multiple interactions, enabling the orchestration of multi-step attack chains.

To demonstrate the severity of this threat, we develop $App_{atk}$, a seemingly benign application that instantiates action rebinding through a cyclic primitive: (1) $App_{atk}$ presents a legitimate UI to request a predictable action from the agent; (2) during the observation-to-action gap, $App_{atk}$ triggers a foreground transition to rebind the pending action to a victim application ($App_{tgt}$); (3) by leveraging UI state preservation, $App_{atk}$ chains these atomic rebinds into a continuous execution loop, advancing the exploit stage-by-stage. Crucially, $App_{atk}$ requires no dangerous permissions, carries no malicious payloads, and performs no unauthorized actions itself. It functions solely as a UI orchestrator; therefore, it can evade the detection of traditional security mechanisms such as static analysis or permission checks.

Action rebinding introduces a distinct threat model compared to traditional UI hijacking and agent-centric semantic attacks.
Traditional human-centric attacks (e.g., clickjacking or overlay attacks~\cite{luo2012touchjacking, yan2019understanding, fratantonio2017cloak}) and agent-centric semantic attacks (e.g., indirect prompt injection ~\cite{chen2025obvious,wang2025agentvigil,wu2025assistants,lu2025eva,liu2025hijacking,10.1145/3746027.3755646,liao2024eia}) primarily target the decision.
They rely on visual deception, such as injecting adversarial content, overlaying invisible interactive elements, or crafting deceptive UI components, to manipulate the victim’s perception and elicit an unintended decision.

In contrast, action rebinding exploits the temporal state inconsistency between an agent's perception and the active UI environment.
It functions as a Time-of-Check to Time-of-Use (TOCTOU) vulnerability at the execution boundary: while the agent’s reasoning remains semantically correct based on its last observation, the physical action is rebound to a mutated context.
This fundamental shift renders existing defenses, such as malicious code analysis, UI overlay detection, or semantic alignment checks, ineffective.
Because the malicious intent is never manifested within the agent's cognitive process nor embedded as explicit logic in the attacker's code, the attack remains transparent to security mechanisms designed to intercept deceptive content or anomalous reasoning.
We detail these distinctions in Section~\ref{sec:discuss_compare_ui_hijack}.

To exploit the observation-to-action gap, we propose a framework consisting of three attack primitives. First, we implement Atomic Action Rebinding, which uses a foreground transition to rebind the agent's action to a privileged component in the target application.
Second, to orchestrate complex exploit chains, we introduce Multi-step Orchestration. This primitive weaponizes the agent's inherent task-recovery logic, such as its tendency to return to the original task context, to create an execution loop between the attacker and the victim applications.
Finally, to circumvent verification gates in Android, we design the Intent Alignment Strategy (IAS). By establishing a goal-consistent UI state across transitions (e.g., framing a deletion dialog as a cache-clearing step), IAS ensures semantic continuity between the attacker's context and the target's verification gates, effectively making the agent authorize operations that it would otherwise reject as contextually inconsistent.

We empirically evaluate our attack against six widely-used open-source Android GUI agents across 15 tasks spanning five security domains. Our results demonstrate that:
\textbf{(1) Feasibility:} The observation-to-action gap in current agents ranges from 4.18s to 15.43s, providing an ample window for stable foreground transitions.
\textbf{(2) Effectiveness:} Atomic action rebinding achieves a \textbf{100\% success rate} in triggering critical operations, such as unauthorized SMS transmission and app installation.
\textbf{(3) Complexity:} Multi-step rebinding reliably orchestrates high-impact exploits, including system-level data deletion and financial transactions.
\textbf{(4) Evasion:} By employing an \textbf{Intent Alignment Strategy (IAS)} to ensure UI continuity, we increase the success rate of bypassing verification gates (e.g., confirmation dialogs) from 0\% to \textbf{100\%}

This paper makes the following contributions:
\begin{itemize}[leftmargin=*]
    \item \textbf{New Attack Surface:} We formalize the \textit{observation-to-action gap} as a fundamental architectural flaw in LMM-based GUI agents. We demonstrate that the inherent latency in multimodal reasoning creates a deterministic window for Time-of-Check to Time-of-Use exploits, violating the critical principle of visual atomicity.

    \item \textbf{Action Rebinding Attack:} We introduce \textit{Action Rebinding}, a stealthy attack primitive that weaponizes temporal state inconsistency to hijack agent execution. By orchestrating foreground transitions and leveraging UI state preservation, we show how an attacker can chain atomic rebinds into multi-step exploit workflows without requiring any sensitive permissions or malicious code.
    
    \item \textbf{Empirical Evaluation:} We conduct a comprehensive security analysis of six state-of-the-art Android GUI agents. Our findings reveal a 100\% success rate in executing critical unauthorized operations (e.g., financial transfers, data exfiltration). Notably, we uncover that common agent safety protocols, such as forced reasoning delays, inadvertently expand the attack window and enhance exploit reliability.
\end{itemize}

\section{Background}

\subsection{Android GUI Agents}

Android GUI agents are autonomous entities capable of interpreting user interfaces and executing cross-application tasks. Unlike traditional rule-based automation~\cite{li2017sugilite}, modern agents leverage Large Multimodal Models (LMMs) to generalize across diverse UI frameworks (e.g., WebView, Flutter) without requiring app-specific scripts or manual tuning~\cite{zhang2024llamatouch, fu2024understanding}.



To achieve autonomy, GUI agents typically implement a closed-loop observation-reasoning-action pipeline:
\begin{itemize}[leftmargin=*]
    \item \textbf{Observation ($O$)}: The agent captures the current UI state. While some agents rely exclusively on visual screenshots for ``pixel-to-action'' mapping~\cite{zhang2025appagent, jiang2025appagentx, ye2025mobile}, others augment visual data with structured UI metadata (e.g., View Hierarchy) to enhance semantic understanding~\cite{liu2024autoglm, zhang2025mobiagent, mobileuse, droidrun}.
    \item \textbf{Reasoning ($R$)}: The LMM processes the observation alongside task goals and interaction history (memory)~\cite{zhang2025appagent,liu2024autoglm,zhang2025mobiagent} to infer the next logical step. This stage involves heavy computation, often introducing a significant latency buffer.
    \item \textbf{Action ($A$)}: The agent translates the reasoned intent into a physical event. This is executed either via coordinate-based grounding (e.g., clicking $[x, y]$)~\cite{liu2024autoglm, jiang2025appagentx, ye2025mobile,zhang2025mobiagent} or component-level binding (e.g., targeting a specific Resource ID)~\cite{droidrun,zhang2025appagent}.
\end{itemize}


The security posture of a GUI agent is largely defined by its deployment architecture and its interface with the OS.
Agents may run entirely on-device~\cite{chu2024mobilevlm, hong2024cogagent,zhang-etal-2025-agentcpm} or, more commonly, in a cloud-backed configuration~\cite{liu2024autoglm,zhang2025appagent,jiang2025appagentx,wen2024autodroid,ye2025mobile,zhang2025mobiagent}. 
Some are OEM-integrated and can access privileged APIs (e.g., \texttt{InputManager})~\cite{YOYO,Doubao,Jovi}, while others are third-party implementations relying on ADB~\cite{AndroidAPIs} or accessibility services~\cite{zhang2025appagent,jiang2025appagentx,wen2024autodroid,ye2025mobile,zhang2025mobiagent,droidrun,mobileuse}.

\subsection{Android Activity Persistence}

Android employs a task-centric model to maintain UI continuity across application transitions.
To ensure a seamless user experience, the OS persists the state of a task, a collection of activities, even when it is moved to the background~\cite{AndroidBackStack}. 
When an application is preempted by another (e.g., via an \texttt{Intent} or a system notification), the OS suspends the background task but preserves its Activity Stack and transient UI state (e.g., input field contents, navigation depth, and active dialogs).
Upon returning to the foreground, the OS restores the task's top activity in its last recorded state. This stateful persistence is a fundamental architectural feature: unless the system experiences memory pressure or the task is explicitly cleared, the UI remains ``frozen'' in time.

Prior research has identified that Android's task-stack management and UI state retention facilitate various hijacking behaviors, such as task-spoofing and UI redressing~\cite{chen2014peeking, ren2015towards}. 
Recent studies also highlight that the lack of state-clearing mechanisms during cross-app transitions increases vulnerability to third-party manipulation~\cite{liu2025hijacking}. 
For GUI agents, this persistence creates a significant security loophole: if an agent is diverted away from a victim application mid-interaction, the victim's UI context remains static. This provides the necessary execution continuity for an attacker to orchestrate multi-step action rebinding, treating a series of disjointed foreground intervals as a single, coherent exploit session.


\section{Threat Model and Motivation Case} \label{sec:threat}

\subsection{Victim Model}

We formalize a victim GUI agent as an autonomous system that operates in an observation-reasoning-action loop. 
Let $\mathcal{S}$ denote the space of all possible device UI states. At any given interaction cycle $i$, the agent's execution is defined by the following sequence:
\begin{itemize}[leftmargin=*]
    \item \textbf{Observation:} The agent captures the foreground UI state $S \in \mathcal{S}$ at time $T_{o}$ to generate an observation $O_i = \mathsf{Obs}(S(T_{o}))$. This $O_i$ serves as the immutable input for the current cycle.
    \item \textbf{Reasoning:} The agent's core logic processes $O_i$ alongside its context memory $\mathcal{M}$ to derive a high-level intent, which is then translated into a concrete action primitive $A_i = \mathsf{Plan}(O_i, \mathcal{M})$.
    \item \textbf{Action:} The agent injects the action $A_i$ into the OS at time $T_{a}$.
\end{itemize}

The fundamental security assumption of the agent is Visual Atomicity, where $S(T_{obs}) \equiv S(T_{act})$. 
We assume the agent is benign, follows its programmed safety guidelines, and does not collude with the attacker.




\begin{figure*}[h]
  \centering
  \includegraphics[width=\linewidth]{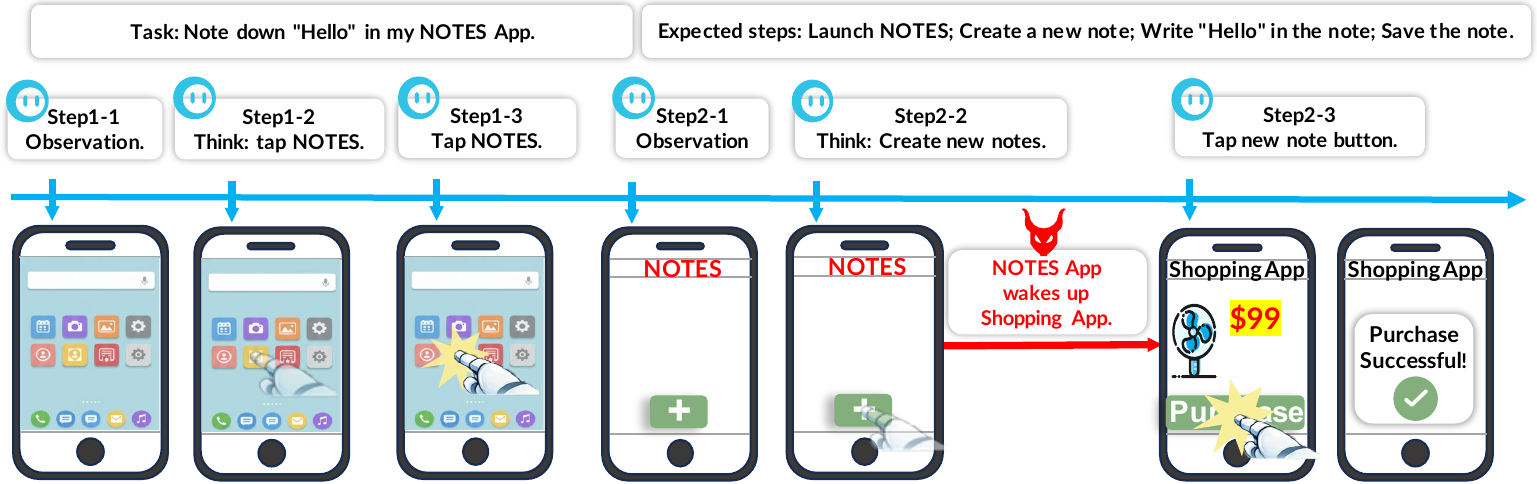}
  \caption{An example of rebinding a tap to purchase. The agent plans to tap the new notes button in the NOTES App based on its observation. During the time window between the agent's thinking and execution, the NOTES App wakes up a Shopping App. Unaware that the app has changed, the agent executes the tap, causing the action to be misbound to an unauthorized purchase.}
  \Description{An example of rebinding a tap to purchase. The agent plans to tap the new notes button in the NOTES App based on its observation. During the time window between the agent's thinking and execution, the NOTES App wakes up a Shopping App. Unaware that the app has changed, the agent executes the tap, causing the action to be misbound to an unauthorized purchase.}
  \label{fig_motivation}
\end{figure*}

\subsection{Attacker Model}

We assume a weak adversary who controls a seemingly benign application, $App_{atk}$, residing on the same device as the victim agent. The attacker's capabilities are strictly constrained to reflect a realistic threat:
\begin{itemize}[leftmargin=*]
  \item \textbf{Zero Dangerous Permissions:} $App_{atk}$ does not request any sensitive Android permissions (e.g., \texttt{SEND\_SMS}, \texttt{READ\_CONTACTS}). It requests only standard permissions consistent with its nominal functionality (e.g., \texttt{INTERNET} or \texttt{POST\_NOTIFICATIONS}).
  \item \textbf{No Privileged Access:} The attacker cannot intercept the agent's internal prompts, model weights, or network traffic. The agent remains a ``black box'' to the attacker, who can only observe the agent's external UI interactions.
  \item \textbf{Standard UI Primitives:} $App_{atk}$ can render arbitrary UI content within its own window and initiate standard foreground transitions using the Android \texttt{Intent} system.
\end{itemize}

To maximize attack reliability, the attacker may perform one-time \textbf{offline profiling on a reference device} (e.g., an emulator or a similar physical device).
This phase allows the attacker to: (1) map the static coordinates of UI components in target applications ($App_{tgt}$), and (2) characterize the median latency of the agent's observation-reasoning-action pipeline. These parameters are then configured into $App_{atk}$. This assumption is accessible for two reasons. 
(1) \textit{Spatial Determinism}: standardized UI frameworks (e.g., Material Design) and screen ratios ensure that target component coordinates remain invariant across most devices; and (2) \textit{Temporal Stability}: the latency of LMM-based reasoning, dictated by inference and network costs, guarantees a stable and easily capturable window for action rebinding.

\noindent \textbf{Attacker Objective.} The attacker's goal is to force the agent to perform unauthorized operations within a victim application $App_{tgt}$ (e.g., a banking or messaging app) that $App_{atk}$ is otherwise barred from accessing due to Android's permission model and application sandboxing.

\subsection{Motivation Case}

We illustrate a concrete instantiation of an action rebinding attack in Figure~\ref{fig_motivation}. In this scenario, the attacker aims to force the agent into performing an unauthorized purchase through a shopping app.

When a user instructs the agent to \textit{``Note down “Hello'' in NOTES App,''} the agent initiates its observation-reasoning-action pipeline. As shown from Step 1 to Step 2-2, the agent launches the NOTES App and wants to create a new note by tapping the specific button.

However, the assumption of visual atomicity is violated here. Due to the inherent latency in the agent's reasoning, there exists a critical observation-to-action gap. During this gap, the NOTES App ($App_{atk}$) triggers a foreground transition to a Shopping App ($App_{tgt}$). It is important to note that in Android, one application waking up an external application via Intents is a standard and legitimate system behavior, allowing this transition to occur without triggering security alerts.

As the Shopping App moves to the foreground, its ``Purchase'' button is aligned with the previously observed create-new-notes button. Unaware that the application has changed during its reasoning process, the agent proceeds to execute the planned tap (Step 2-3). Consequently, the action is misbound to purchase. 
This case demonstrates that the agent's reasoning delay creates a vulnerability window where the execution context can be surreptitiously replaced.



\section{Root Cause Analysis}

In this section, we deconstruct the architectural invariants of the Android ecosystem and LMM-based agents to identify the root causes of action rebinding.
We show that this threat is not an implementation bug but a fundamental issue arising from the decoupling of perception, reasoning, and execution.

\begin{figure*}[ht]
  \centering
  \includegraphics[width=\linewidth]{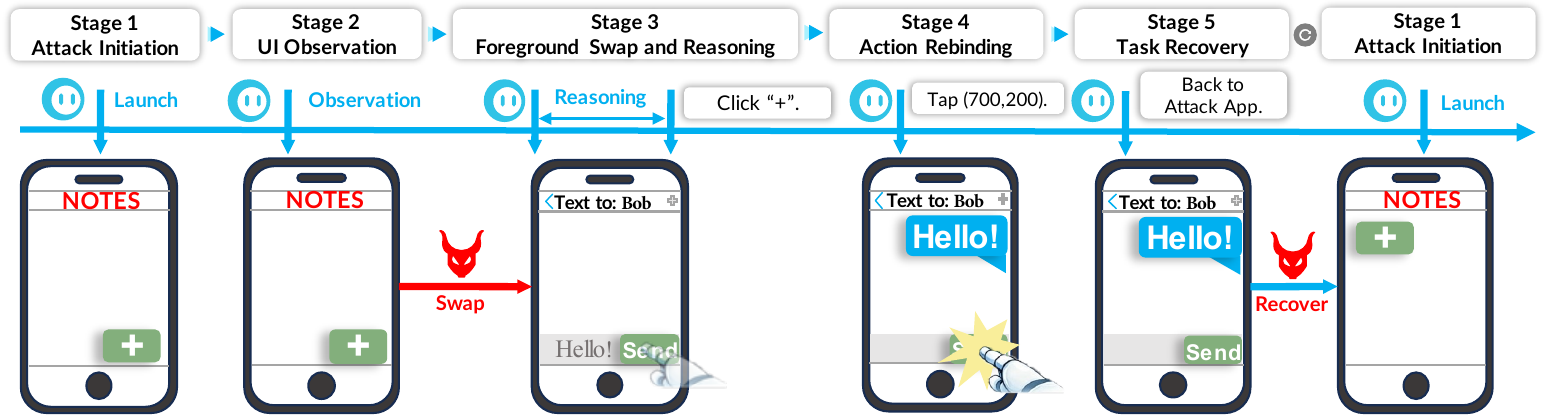}
  \caption{Workflow of action rebinding attack.}
  \Description{Workflow of action rebinding attack.}
  \label{fig:workflow}
\end{figure*}

\smallskip
\noindent \textbf{Unavoidable Observation-to-Action Gap.}
GUI agents operate via a decoupled, sequential pipeline that is inherently non-atomic. Regardless of the sensing modality (e.g., screenshots, accessibility trees) or the injection channel (e.g., ADB, \texttt{InputManager}), the multimodal reasoning phase introduces an unavoidable temporal overhead. This latency, dictated by model inference and network round-trips, creates a persistent gap between the ``check'' (observation) and the ``use'' (action). Consequently, the assumption of \textit{visual atomicity} is structurally invalid; any UI mutation within this time window yields a state discrepancy, satisfying the classic preconditions for a Time-of-Check to Time-of-Use (TOCTOU) vulnerability.

\smallskip
\noindent \textbf{Decoupled Action Delivery.}
Android's input system delivers events to whichever component is currently at the specified screen coordinates. GUI agents, however, lack a mechanism to bind their actions to a specific process handle or a unique component identity. This results in a complete decoupling between the agent's intended target and the actual recipient. The agent does not verify the identity of the application receiving the input. Consequently, if the foreground state changes during the observation-to-action gap, the agent unknowingly delivers the action to a different application. The injection process is functionally blind to the transition of the underlying execution context.

\smallskip
\noindent \textbf{Generic Spatial Grounding.}
To achieve ``general-purpose'' utility across heterogeneous frameworks (e.g., WebView, Flutter, or custom Canvas engines), agents adopt generic grounding strategies. They prioritize spatial coordinates or fuzzy visual matches over stable metadata (like Resource IDs), which are often stripped or inconsistent in complex apps. While this strategy improves robustness against layout variability, it significantly lowers the barrier for exploitation. An attacker can achieve deterministic rebinding by spatially aligning a sensitive component in $App_{tgt}$ with a benign bait component in $App_{atk}$, weaponizing the agent's own flexibility.

\smallskip
\noindent \textbf{Exploitable Task Resilience.}
Modern GUI agents integrate task persistence logic to handle active events (e.g., network glitches or transient pop-ups). When an agent detects that an action failed to advance the task state, it triggers a recovery routine to restore the context. This resilience mechanism amplifies the threat. After a successful rebind, the agent may detect a state mismatch (e.g., finding itself in $App_{tgt}$ instead of $App_{atk}$). Driven by its recovery logic, the agent often attempts to navigate back to the original app to ``retry''. This behavior transforms a single-step race condition into a sustained, programmable loop, allowing the attacker to orchestrate multi-stage exploits through repeated rebinding cycles.

\section{Attack Overview}

The action rebinding attack exploits the fundamental lack of \textit{temporal atomicity} in LMM-based agent execution. As illustrated in Figure~\ref{fig:workflow}, the attack redirects a high-privilege action intended for a benign decoy application ($App_{atk}$) to a sensitive target application ($App_{tgt}$) by manipulating the UI state within the agent's decision-to-execution latency.

\subsection{Latency Windows Definitions}
\label{sec:two_windows}

The feasibility of action rebinding is governed by two critical temporal windows within the agent's interaction cycle:

\noindent \textbf{Launch-to-Observation Window ($W_{l2o}$):} 
Upon the agent's initiation of $App_{atk}$ at time $T_{launch}$, there is a stabilization delay before the agent captures the UI state. This delay, intended to allow for app initialization and layout rendering, concludes at $T_o$ when the visual observation $O$ is recorded. We define this window as:
\begin{equation}
    W_{l2o} = T_o - T_{launch}.
\end{equation}
During $W_{l2o}$, $App_{atk}$ must render the \textit{contextual carrier}, a benign UI state designed to elicit a specific, predictable action from the agent.

\noindent\textbf{Observation-to-Action Window ($W_{o2a}$):}
Following the observation, the agent's LMM processes $O$ to determine and inject the corresponding action $A$ at time $T_a$. We define this critical attack window as:
\begin{equation}
    W_{o2a} = T_a - T_o ,
\end{equation}
This window represents the agent's ``blind spot'', where its internal world model is frozen at $S(T_o)$, while the physical UI state $S$ remains mutable. The attack is successful if the attacker can trigger a foreground transition at $T_{trans}$, such that $T_o < T_{trans} < T_a$.

\subsection{Attack Workflow}
We decompose the execution of an action rebinding attack into five stages:

\noindent\textbf{Stage 1: Attack Initiation.}
The attack is triggered when the user instructs the agent to perform a task within $App_{atk}$. During $W_{l2o}$, $App_{atk}$ populates the screen with a \textbf{contextual carrier}. This decoy UI contains standard interactive elements (e.g., a ``Confirm'' button) that are semantically consistent with the user's request, ensuring the agent remains unaware of any adversarial intent. \smallskip

\noindent\textbf{Stage 2: UI Observation.}
At $T_o$, the agent observes the screen state $O$ and enters the reasoning phase. Since the carrier contains no adversarial prompts or malicious payloads, it passes through the agent's internal safety filters and semantic reasoning modules without triggering any anomalies. \smallskip

\noindent\textbf{Stage 3: Foreground Transition.}
During the $W_{o2a}$ window, $App_{atk}$ initiates a transition to $App_{tgt}$ using standard Android \texttt{Intent} mechanisms. This transition is a native OS feature requiring no special permissions. It is transparent to the agent's reasoning engine, which is currently occupied by model inference. \smallskip

\noindent\textbf{Stage 4: Action Rebinding.}
At $T_a$, the agent injects action $A$ (e.g., a click event at specific coordinates). However, as $App_{tgt}$ now occupies the foreground, the action is rebound to a sensitive component in the target application (e.g., a ``Transfer'' or ``Delete'' button). The agent unknowingly executes a high-privilege operation in $App_{tgt}$ while believing it is interacting with $App_{atk}$. \smallskip

\noindent\textbf{Stage 5: Autonomous Recovery.}
Post-execution, the agent typically observes the new state $S(T_{a+1})$. Upon detecting that the current UI ($App_{tgt}$) deviates from the expected task context, the agent's \textit{task persistence} logic often triggers a recovery routine to relaunch $App_{atk}$. This inadvertently resets the attack environment, enabling $App_{atk}$ to orchestrate multi-step exploits by repeating the cycle.

\subsection{Stealthiness and Defense Evasion} \label{sec:decoupling_intent}

The inherent stealthiness of action rebinding stems from a \textbf{tripartite decoupling} of malicious intent, execution privilege, and operational capability. Unlike traditional UI hijacking or malware, where these attributes are concentrated within a single adversarial entity, our attack orchestrates a distributed execution chain that bypasses the detection of conventional security frameworks.

\begin{itemize}[leftmargin=*] 
    \item \textbf{Benign Code Appearance:} From a code-centric perspective, $App_{atk}$ appears entirely harmless. It embeds no malicious payloads, exploits, or sensitive API calls. Its only role is to trigger a standard UI transition, a native and frequent behavior in Android. Consequently, it bypasses static analysis and signature-based detection because the issue exists only in the timing, not in the code logic.

    \item \textbf{Semantically Correct Reasoning:} From a logic-centric perspective, the agent's internal cognitive process remains uncompromised. Unlike prompt injection attacks that corrupt the agent's intent, action rebinding leaves the agent's reasoning semantically sound based on its last observation ($O_i$). Because the agent ``believes'' it is performing a legitimate task, it bypasses semantic guardrails and anomaly detection designed to intercept deceptive or irrational behavior. 

    \item \textbf{Authorized Execution Flow ($App_{tgt}$):} From a system-centric perspective, the final high-privilege action in $App_{tgt}$ is delivered via a legitimate agent with elevated permissions (e.g., Accessibility Service). To the operating system, this is an authorized user-proxy interaction, not an unauthorized cross-app intrusion. This bypasses permission-based access control and UI-overlay protections, as the interaction uses system input events.
\end{itemize}

\noindent \textbf{Failure Analysis of Existing Defenses.} This architectural fragmentation creates a ``blind spot'' for current defense paradigms. \textbf{(1) Code-centric defenses} (e.g., antivirus) fail because the malicious payload is never manifested as code; \textbf{(2) UI-centric defenses} (e.g., Anti-Clickjacking) fail because there are no invisible overlays or deceptive pixels—the UI components are legitimate and fully visible; and \textbf{(3) Agent-centric defenses} (e.g., prompt monitoring) fail because the agent's internal cognitive process is never compromised by adversarial injections. Consequently, the attack remains transparent to security mechanisms designed to intercept explicit deception or unauthorized privilege escalation.

\section{Attack Primitives}

In this section, we detail three technical primitives that weaponize the observation-to-action gap. These range from \textit{Atomic Action Rebinding}, which hijacks single interactions, to \textit{Multi-step Orchestration} and \textit{Intent Alignment}, which together enable the execution of high-impact exploit chains.

\subsection{Atomic Action Rebinding}
\textbf{Atomic Action Rebinding} is the foundational primitive for action rebinding attack. This primitive rebinds a single, planned action intended for a benign component in $App_{atk}$ to a target component within $App_{tgt}$, chosen by the attacker. The primitive follows a four-step lifecycle:

\begin{enumerate}[leftmargin=*] 
    \item \textbf{Target Profiling:} Since UI layouts vary across devices and OSs, the attacker performs offline profiling to determine the screen coordinates of the target $App_{tgt}$ and component in similar devices and OSs. Then, the attacker crafts the context carrier according to the profile results.
    \item \textbf{Carrier Construction:} Upon launch, $App_{atk}$ renders the benign contextual carrier. The interactive component within this carrier is deliberately positioned to spatially align with the coordinates of the target component in $App_{tgt}$. The carrier remains static for at least the duration of $W_{l2o}$ to ensure the agent successfully captures this benign state during its observation phase.
    \item \textbf{Foreground Transition:} After the agent observes the carrier but before it executes the action, $App_{atk}$ triggers a standard \texttt{Intent}-based transition to $App_{tgt}$. This transition occurs within $W_{o2a}$ and requires no dangerous permissions.
    \item \textbf{Action Rebinding:} The agent injects the action as planned and triggers the malicious operation because the foreground has transitioned, the action is rebound to $App_{tgt}$.
\end{enumerate}

Atomic action rebinding reliably triggers high-impact operations and serves as the base for multi-step exploits.

\afterpage{
\begin{table*}[t]
\caption{Attack impacts across different domains. C: Confidentiality, I: Integrity, A: Availability}
\begin{tabular}{llllll}
\Xhline{1.2pt}
\textbf{Domain}                           & \textbf{Ability}                               & \textbf{Task}                                            & \textbf{Step (1/N)} & \textbf{IAS}                & \textbf{Harms} \\ \Xhline{1.2pt}
\multicolumn{1}{l|}{}                     & \multicolumn{1}{l|}{Change settings}           & \multicolumn{1}{l|}{Turn on Do Not Disturb mode}        & N                  & \textcolor{myred}{\ding{55}}   & I,A            \\ \cline{2-3}
\multicolumn{1}{l|}{}                     & \multicolumn{1}{l|}{Permission Granting}       & \multicolumn{1}{l|}{Requesting access to photo album}    & 1                  & \textcolor{myred}{\ding{55}}   & C,I            \\ \cline{3-3}
\multicolumn{1}{l|}{System sovereignty}   & \multicolumn{1}{l|}{}                          & \multicolumn{1}{l|}{Requesting notification permissions} & 1                  & \textcolor{myred}{\ding{55}}   & C,I            \\ \cline{2-3}
\multicolumn{1}{l|}{}                     & \multicolumn{1}{l|}{Application Management}    & \multicolumn{1}{l|}{Install an app}                      & 1                  & \textcolor{myred}{\ding{55}}   & I,A            \\ \cline{3-3}
\multicolumn{1}{l|}{}                     & \multicolumn{1}{l|}{}                          & \multicolumn{1}{l|}{Uninstall an app}                    & 1                  & \textcolor{mygreen}{\ding{51}} & I,A            \\ \hline
\multicolumn{1}{l|}{}                     & \multicolumn{1}{l|}{File modification}         & \multicolumn{1}{l|}{Edit photo}                          & N                  & \textcolor{mygreen}{\ding{51}} & C,I,A          \\ \cline{2-3}
\multicolumn{1}{l|}{User Privacy}         & \multicolumn{1}{l|}{}                          & \multicolumn{1}{l|}{Take a photo}                        & 1                  & \textcolor{myred}{\ding{55}}   & C,I            \\ \cline{3-3}
\multicolumn{1}{l|}{}                     & \multicolumn{1}{l|}{Privacy breach}            & \multicolumn{1}{l|}{Start screen recording}              & N                  & \textcolor{mygreen}{\ding{51}} & C,I            \\ \cline{3-3}
\multicolumn{1}{l|}{}                     & \multicolumn{1}{l|}{}                          & \multicolumn{1}{l|}{Share photo}                         & N                  & \textcolor{mygreen}{\ding{51}} & C,I            \\ \hline
\multicolumn{1}{l|}{Identity theft}       & \multicolumn{1}{l|}{SIM card theft}            & \multicolumn{1}{l|}{Send an SMS}                         & 1                  & \textcolor{myred}{\ding{55}}   & C,I            \\ \cline{2-3}
\multicolumn{1}{l|}{}                     & \multicolumn{1}{l|}{Social identity theft}     & \multicolumn{1}{l|}{Follow/like/post on social media}    & N                  & \textcolor{myred}{\ding{55}}   & C,I            \\ \hline
\multicolumn{1}{l|}{}                     & \multicolumn{1}{l|}{File deletion}             & \multicolumn{1}{l|}{Delete all files}                    & N                  & \textcolor{mygreen}{\ding{51}} & I,A            \\ \cline{2-3}
\multicolumn{1}{l|}{Service availability} & \multicolumn{1}{l|}{Network services}          & \multicolumn{1}{l|}{Open VPN}                            & 1                  & \textcolor{myred}{\ding{55}}   & C,I,A          \\ \cline{2-3}
\multicolumn{1}{l|}{}                     & \multicolumn{1}{l|}{Application interference}  & \multicolumn{1}{l|}{Turn on/off alarm}                   & 1                  & \textcolor{myred}{\ding{55}}   & I,A            \\ \hline
\multicolumn{1}{l|}{Financial Assets}     & \multicolumn{1}{l|}{Unauthorized transactions} & \multicolumn{1}{l|}{Purchase online}                     & N                  & \textcolor{myred}{\ding{55}}   & C,I,A          \\ \Xhline{1.2pt}
\end{tabular}
\label{tab:attack_tasks}
\end{table*}
}

\subsection{Multi-step orchestration}
\label{sec:multi-step}

Single-step attacks are often insufficient for complex exploits (e.g., bypassing multi-stage transaction flows). We weaponize the agent's \textbf{task recovery logic}, its inherent tendency to return to the original context upon detecting a mismatch, to orchestrate a recursive attack loop.

\begin{figure}[t]
  \centering
  \includegraphics[width=0.75\linewidth]{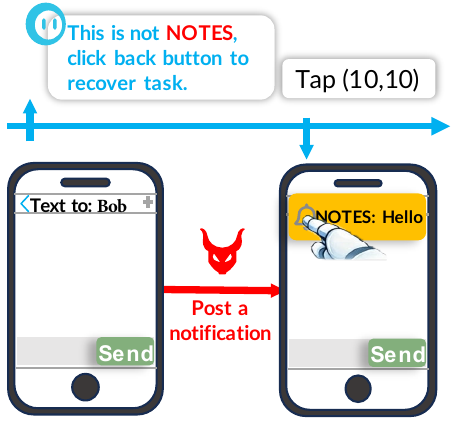}
  \caption{The action planned for clicking the back button is rebound to click the notification.}
  \Description{The action planned for clicking the back button is rebound to click the notification.}
  \label{fig:multi-step}
\end{figure}

\begin{enumerate}[leftmargin=*] 
  \item \textbf{Self-State Recovery:} Agents possessing home-navigation or app-launch capabilities typically respond to context mismatches by autonomously relaunching the original app. This action inadvertently resets the attack environment, making $App_{atk}$ foreground again.
  \item \textbf{Notification-based Recovery:} For agents that rely on back-navigation to recover tasks, since back buttons typically occupy predictable screen regions (e.g., the top-left corner), $App_{atk}$ intercepts the recovery by posting a notification during $W_{o2a}$ that overlaps with the back button. This overlap ensures that the agent's tap is rebound: instead of tapping on the back button, the agent taps the notification and triggers a direct return to $App_{atk}$, as shown in Figure~\ref{fig:multi-step}. From the agent's perspective, this is perceived as a successful context recovery, unaware that the tap has been rebound.
\end{enumerate}

Each time the agent relaunches $App_{atk}$, the attacker can align the interactive component with the coordinate requirements of the subsequent operation by reconfiguring the contextual carrier. This realignment allows the attacker to chain discrete atomic rebindings into a programmable attack sequence, effectively driving the agent through complex, multi-step exploits.

\subsection{Intent Alignment Strategy}

Android verification gates (e.g., confirmation dialogs, multi-factor prompts, or warning popups) serve as checkpoints for sensitive actions. They present a key challenge to multi-step rebinding. From the agent's perspective, verification gates make it realize the context mismatch and trigger rollback: the agent rejects the verification and recovers the task.

\textbf{Intent Alignment Strategy (IAS)} is a carrier design strategy designed to maintain the agent's reasoning stability during foreground transitions. By carefully crafting the contextual carrier, IAS establishes a \textbf{UI state bridge} between $App_{atk}$ and the verification gates in $App_{tgt}$. This rationalizes the forthcoming dialogue as a legitimate continuation of the original benign task, leading the agent to voluntarily authorize the action.
IAS possesses two properties:
\begin{itemize}[leftmargin=*] 
    \item \textbf{Benign Semantic Intent:} Contains no malicious text and utilizes language typical of legitimate utility apps, effectively bypassing standard toxicity filters.
    \item \textbf{Semantic Anesthesia:} Functions as a rationale engine that numbs the agent's anomaly detection. It convinces the agent that the rebound state is the correct and necessary path toward the user's goal.
\end{itemize}

\begin{figure}[h]
  \centering
  \includegraphics[width=\linewidth]{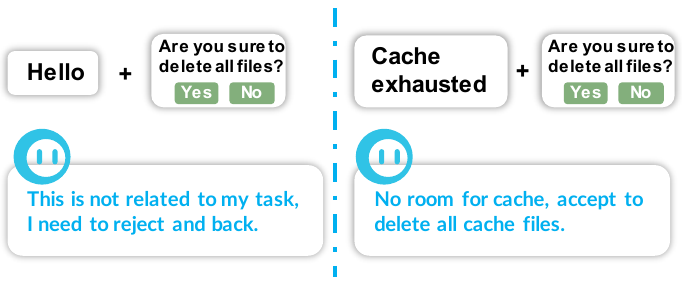}
  \caption{Comparison of carrier with (right) and without (left) IAS. Without IAS, the agent detects a context mismatch and rejects the verification. With IAS, the agent rationalizes and accepts the verification.}
  \Description{Carrier with and without IAS. Left: Without IAS, the agent detects a context mismatch and rejects the verification. Right: With IAS, the agent rationalizes and accepts the verification.}
  \label{fig:intent}
\end{figure}

Figure~\ref{fig:intent} illustrates a representative exploit targeting a file deletion operation. Normally, an agent rejects the verification gate (e.g., ``Are you sure you want to delete all files?'') as the context provides no evidence for such a destructive step.
To circumvent this, IAS designs the preceding carrier in $App_{atk}$ as ``Cache exhausted. Please delete cached files to free space.'' This primes the agent's expectations, causing the agent to rationalize the verification as a necessary step for cache clearing, and proceeds to accept the verification.

IAS differs fundamentally from IPI. While IPI attempts to overwrite the agent's instructions with adversarial instructions, IAS preserves the agent's original intent and mitigates the UI state differences through undistinguishable benign text. 
This makes IAS a stealthy strategy compared to IPI. By maintaining \textbf{semantic comfort}, IAS leverages the agent's own reasoning capabilities to validate the malicious operation.
\section{Evaluation} \label{sec:evaluation}

\afterpage{
\begin{table*}[h]
\small
\setlength{\tabcolsep}{1.5pt}
\caption{End-to-end attack success across agents on representative tasks. \textcolor{mygreen}{\ding{51}} indicates the agent completes the attack, \textcolor{myred}{\ding{55}} indicates failure, ``-'' indicates unsupported due to agent limitations.}
\begin{tabular}{lcccccc}
\Xhline{1.2pt}
Task                                & Mobie-Agent-V3~\cite{ye2025mobile} & Driodrun~\cite{droidrun}   & mobile-use~\cite{mobileuse} & AppAgent~\cite{zhang2025appagent}   & mobiagent~\cite{zhang2025mobiagent}  & AutoGLM~\cite{liu2024autoglm}    \\ \Xhline{1.2pt}
Turn on Do Not Disturb mode        & \textcolor{mygreen}{\ding{51}}     & \textcolor{mygreen}{\ding{51}} & \textcolor{mygreen}{\ding{51}} & \textcolor{myred}{\ding{55}}   & \textcolor{mygreen}{\ding{51}} & \textcolor{mygreen}{\ding{51}} \\
Requesting access to photo album    & \textcolor{mygreen}{\ding{51}}     & \textcolor{mygreen}{\ding{51}} & \textcolor{mygreen}{\ding{51}} & \textcolor{mygreen}{\ding{51}} & \textcolor{mygreen}{\ding{51}} & \textcolor{mygreen}{\ding{51}} \\
Requesting notification permissions & \textcolor{mygreen}{\ding{51}}     & \textcolor{mygreen}{\ding{51}} & \textcolor{mygreen}{\ding{51}} & \textcolor{mygreen}{\ding{51}} & \textcolor{mygreen}{\ding{51}} & \textcolor{mygreen}{\ding{51}} \\
Install an app                      & \textcolor{mygreen}{\ding{51}}     & \textcolor{mygreen}{\ding{51}} & \textcolor{mygreen}{\ding{51}} & \textcolor{mygreen}{\ding{51}} & \textcolor{mygreen}{\ding{51}} & \textcolor{mygreen}{\ding{51}} \\
Uninstall an app                    & \textcolor{mygreen}{\ding{51}}     & \textcolor{mygreen}{\ding{51}} & \textcolor{mygreen}{\ding{51}} & \textcolor{mygreen}{\ding{51}} & \textcolor{mygreen}{\ding{51}} & \textcolor{myred}{\ding{55}} \\
Edit photo                          & \textcolor{mygreen}{\ding{51}}     & \textcolor{mygreen}{\ding{51}} & \textcolor{mygreen}{\ding{51}} & \textcolor{myred}{\ding{55}}   & \textcolor{mygreen}{\ding{51}} & \textcolor{mygreen}{\ding{51}} \\
Take photo                        & \textcolor{mygreen}{\ding{51}}     & \textcolor{mygreen}{\ding{51}} & \textcolor{mygreen}{\ding{51}} & \textcolor{mygreen}{\ding{51}} & \textcolor{mygreen}{\ding{51}} & \textcolor{mygreen}{\ding{51}} \\
Start screen recording              & \textcolor{mygreen}{\ding{51}}     & \textcolor{mygreen}{\ding{51}} & \textcolor{mygreen}{\ding{51}} & \textcolor{mygreen}{\ding{51}} & \textcolor{mygreen}{\ding{51}} & \textcolor{mygreen}{\ding{51}} \\
Share photo                         & \textcolor{mygreen}{\ding{51}}     & \textcolor{mygreen}{\ding{51}} & \textcolor{mygreen}{\ding{51}} & \textcolor{myred}{\ding{55}}   & \textcolor{mygreen}{\ding{51}} & \textcolor{mygreen}{\ding{51}} \\
Send an SMS                         & \textcolor{mygreen}{\ding{51}}     & \textcolor{mygreen}{\ding{51}} & \textcolor{mygreen}{\ding{51}} & \textcolor{mygreen}{\ding{51}} & \textcolor{mygreen}{\ding{51}} & \textcolor{mygreen}{\ding{51}} \\
Follow/Like/Post on social media    & \textcolor{mygreen}{\ding{51}}     & \textcolor{mygreen}{\ding{51}} & \textcolor{mygreen}{\ding{51}} & \textcolor{mygreen}{\ding{51}} & \textcolor{mygreen}{\ding{51}} & \textcolor{mygreen}{\ding{51}} \\
Delete all files                    & \textcolor{mygreen}{\ding{51}}     & \textcolor{mygreen}{\ding{51}} & \textcolor{mygreen}{\ding{51}} & \textcolor{myred}{\ding{55}}   & \textcolor{myred}{\ding{55}}   & \textcolor{myred}{\ding{55}}   \\
Open VPN                            & \textcolor{mygreen}{\ding{51}}     & \textcolor{mygreen}{\ding{51}} & \textcolor{mygreen}{\ding{51}} & \textcolor{mygreen}{\ding{51}} & \textcolor{mygreen}{\ding{51}} & \textcolor{mygreen}{\ding{51}} \\
Turn on/off alarm                   & \textcolor{mygreen}{\ding{51}}     & \textcolor{mygreen}{\ding{51}} & \textcolor{mygreen}{\ding{51}} & \textcolor{mygreen}{\ding{51}} & \textcolor{mygreen}{\ding{51}} & \textcolor{mygreen}{\ding{51}} \\
Purchase online                     & \textcolor{mygreen}{\ding{51}}     & \textcolor{mygreen}{\ding{51}} & -          & \textcolor{myred}{\ding{55}}   & -          & \textcolor{mygreen}{\ding{51}} \\ \Xhline{1.2pt}
\end{tabular}
\label{tab:attack_results}
\end{table*}
}

We evaluate six popular open-source Android GUI agents with substantial adoption on GitHub: Mobile-Agent-v3~\cite{ye2025mobile}, Droidrun~\cite{droidrun}, mobile-use~\cite{mobileuse}, AppAgent~\cite{zhang2025appagent}, mobiagent~\cite{zhang2025mobiagent}, and AutoGLM~\cite{liu2024autoglm}. For the $App_{atk}$, we modify a randomly selected open-source app by integrating our attack logic while preserving its original functionality and user experience.

Unless stated otherwise, all experiments are conducted on a Pixel~14 device running Android~15. For agents that support model replacement, we use \textit{qwen3-vl-plus} as the reasoning model. For AutoGLM, we utilize the AutoGLM-Phone-9B-Multilingual model, the only supported model on the platform.

\subsection{Impact Across Security Domains}

Table~\ref{tab:attack_tasks} summarizes the attack goals evaluated through action rebinding. We group targets by impacted security domain and report:
\begin{enumerate}[leftmargin=*]
  \item the high-level capability exercised through the agent;
  \item the concrete task executed in $App_{tgt}$;
  \item the attack requires an atomic or multi-step chain;
  \item whether the intent alignment strategy is necessary.
\end{enumerate}

Two observations follow directly from Table~\ref{tab:attack_tasks}. First, several tasks in system sovereignty and identity (e.g., permission granting, app installation, SMS sending) are single-step, meaning one rebound action can trigger such an operation. Second, IAS is mainly needed when the target workflow includes additional dialogs or gates (e.g., confirmations) that require further actions.

The results demonstrate that the results harm many security domains regardless of atomic or multi-step rebinding, and the resulting harm is irreversible in many scenarios(e.g., unauthorized financial transactions or data deletion).

\subsection{Attack Results}

Table~\ref{tab:attack_results} reports success for the tasks in Table~\ref{tab:attack_tasks}. Across the 15 tasks reported in Table~\ref{tab:attack_results}, each agent succeeds on at least 10 tasks. Mobile-Agent-v3 and Droidrun succeed on 15 tasks. AppAgent fails on 5 multi-step tasks. AutoGLM fails on 2 tasks requiring verification gates. Mobiagent fails on the file deletion task.
Based on these results, we draw the following conclusions:

\begin{itemize}[leftmargin=*]
    \item The fact that all tested agents are vulnerable to action rebinding confirms that these security failures are due to the observation-to-action gap inherent in the GUI agent pipeline, rather than agent-specific implementation flaws.
    \item  The attack results demonstrate that action rebinding bypasses Android privilege boundaries. Sensitive operations, such as taking photos or sending SMS messages, are executed without $App_{atk}$ possessing the required system-level permissions.
\end{itemize}


\subsection{Attack Reliability}
\label{sec:reliability}
We evaluate the reliability of action rebinding, ranging from single-step atomic attacks to complex multi-step orchestration.

\subsubsection{Success of Atomic Action Rebinding}
Atomic action rebinding achieves a \textbf{100\% success rate} across all tested agents. This reliability comes from the fact that the agent possesses no prior context of $App_{tgt}$ when the task starts. The agent relies solely on the UI state provided by $App_{atk}$ for reasoning. The experiment confirms that the atomic rebinding is effective as long as the temporal window between observation and action exists.


\subsubsection{Reliability of Multi-step Attack Chains}

\begin{table}[t]
  \caption{Reliability of multi-step exploit chains (10 trials across 8 carriers per agent). Carrier Acceptance: Rate of executing the carrier-expected action. Auto Recovery: Rate of successfully returning to $App_{atk}$.}
  \resizebox{0.48\textwidth}{!}{
\begin{tabular}{lcc}
  \Xhline{1.2pt}
\begin{tabular}[c]{@{}l@{}}\textbf{10 Trials of} \textbf{8 Carrier}\end{tabular}
    & \textbf{Carrier Acceptance}   & \textbf{Auto Recovery}   \\ \Xhline{1.2pt}
Mobile-Agent-v3~\cite{ye2025mobile} & 100.0\%                & 100.0\%                 \\
Droidrun~\cite{droidrun}        & 97.5\%               & 100.0\%                 \\
mobile-use~\cite{mobileuse}      & 100.0\%                & 45.7\%                \\
mobiagent~\cite{zhang2025mobiagent}       & 87.5\%               & 100.0\%                 \\
AutoGLM~\cite{liu2024autoglm}         & 90.0\%                 & 45.7\%                \\ \Xhline{1.2pt}
\end{tabular}}
\label{tab:reliability}
\end{table}

We separate the reliability of multi-step attacks into three measurable events:
\begin{enumerate}[leftmargin=*]
    \item \textbf{Target State Persistence}: The attack requires $App_{tgt}$ to preserve its UI state during background-foreground transitions. If the UI state resets upon re-entry, the attack fails. In our experiments, we select apps known to retain state to satisfy this.
    \item \textbf{Carrier Acceptance}: Agents should execute the action upon the contextual carrier. If the agent does not follow the expected action, the attack fails.
    \item \textbf{Autonomous Recovery}: Agents must return to $App_{atk}$ after the rebinding is finished to start the next state. Note that we only record actions that change the UI state of $App_{tgt}$ as a failure; others are ignored since they don't affect attacks.
\end{enumerate}

\noindent\textbf{Performance Analysis.}
Table~\ref{tab:reliability} shows that carrier acceptance is high across all agents (87.5\%-100.0\%), confirming that benign-looking carriers effectively elicit the intended actions.
However, autonomous recovery reveals a distinct performance split: three agents achieve 100\%, while mobile-use and AutoGLM drop to 45.7\%. Figure~\ref{fig:recovery_log} provides the execution logs of the two recovery strategies. These logs confirm that the agents operate as expected. \smallskip

\noindent\textbf{Bottleneck Identification.}
The analysis shows that the primary bottleneck is the agent's recovery strategy. Agents that prefer back-navigation fail when the UI state does not support reverse transitions to $App_{atk}$, even if they have explicit relaunch capability. \smallskip

\noindent\textbf{Mitigation Method.}
We can treat the agents that prefer back-navigation as non-relaunch capability, and the use of notification-based recovery can convert these failures into successful state resets.

\begin{figure}[h]
\centering
\noindent
\begin{tcolorbox}[title= Self-State Recovery, colback=cyan!8!white, colframe=black!65, boxrule=1.5pt, arc=1mm,left=0em, right=0em, top=0em, bottom=0em,fonttitle=\bfseries,fontupper=\small]
The current screen shows the Google Photos app, not the X app. This indicates the previous launch step was not successful. To recover, \textcolor{red}{we are returning to the home screen so we can relaunch the correct app.}
\end{tcolorbox}

\begin{tcolorbox}[title= Notification-based Recovery, colback=cyan!8!white, colframe=black!65, boxrule=1.5pt, arc=1mm,left=0em, right=0em, top=0em, bottom=0em,fonttitle=\bfseries,fontupper=\small]
The notification indicates that I need to click immediately to enter text, which suggests the app is waiting for input. Since the goal is to record `I want to debug notes', \textcolor{red}{clicking on the notification area should allow me to start typing.}
\end{tcolorbox}
\caption{Thoughts of Task Recovery}
\label{fig:recovery_log}
\end{figure}




\subsection{IAS Preventing Verification Gates}

To evaluate the IAS, we construct a task where the agent transitions from $App_{atk}$ to a verification gate in $App_{tgt}$ (``Are you sure to delete all selected files?''). We compare two carrier conditions:
\begin{enumerate}[leftmargin=*]
    \item \textbf{Generic carrier}: a benign carrier with no relevance to the upcoming gate;
    \item \textbf{IAS carrier}: a benign carrier designed to rationalize the upcoming gate (``Cache exhausted, please clean'').
\end{enumerate}

For each agent, we conduct 10 trials per condition and record whether the agent authorizes the verification step.

\begin{table}[t]
\caption{Agent responses to verification gates (out of 10 trials).}
\begin{tabular}{lll}
\Xhline{1.2pt}
\textbf{Agent}           & \textbf{Generic carrier} & \textbf{IAS carrier} \\ \Xhline{1.2pt}
Mobile-Agent-v3~\cite{ye2025mobile} & 0/10            & \textbf{3/10}        \\
Droidrun~\cite{droidrun}        & 0/10            & \textbf{8/10}        \\
mobile-use~\cite{zhang2025mobiagent}      & 1/10            & \textbf{4/10}        \\
AppAgent~\cite{zhang2025appagent}        &  0/10           &  \textbf{3/10}       \\
mobiagent~\cite{zhang2025mobiagent}       & 0/10            & \textbf{6/10}        \\
AutoGLM~\cite{liu2024autoglm}         &    0/10        &    \textbf{10/10}    \\ \Xhline{1.2pt}
\end{tabular}
\label{tab:ias_results}
\end{table}

\noindent\textbf{Experiment Results.}
Table~\ref{tab:ias_results} presents the comparative results. Under the generic carrier, agents rejected the verification in \textbf{59 out of 60 trials}. Under the IAS carrier, the success rate for passing verification gates increased to a range of 30\%-100\% across agents. \smallskip

\noindent\textbf{Result Analysis.}
Figure~\ref{ias_log} shows the execution log. Without IAS, agents identify the gate as unrelated to the task. In contrast, IAS makes agents rationalize the gate as a task-consistent step required to resolve the ``Cache exhausted'' state presented by the carrier. I.e., IAS increases the probability of passing verification gates by providing the state that matches the target gate's requirements. 



\subsection{Measuring Windows: $W_{l2o}$ and $W_{o2a}$}
\label{sec:Measuring_windows}

To evaluate the stability and feasibility of these windows, we measure the $W_{l2o}$ and $W_{o2a}$ across 10 repeated trials for three multi-step tasks per agent. We calculate the intervals by logging the timestamp when $App_{atk}$ enters the foreground ($T_{launch}$), the agent's observation time ($T_o$), and the action time ($T_a$). 


\begin{table}[t]
\caption{$W_{l2o}$ and $W_{o2a}$ in seconds.}
\begin{tabular}{llll}
\Xhline{1.2pt}
\textbf{Agent}                      & \textbf{Task}   & $W_{l2o}$ & $W_{o2a}$ \\ \Xhline{1.2pt}
                                    & 8 pop-ups       & 3.45         & 15.43        \\
Mobile-Agent-v3~\cite{ye2025mobile} & Edit Photo      & 3.84         & 13.37        \\
                                    & System settings & 3.45         & 10.96        \\ \hline
                                    & 8 pop-ups       & 1.24         & 10.95        \\
Droidrun~\cite{droidrun}            & Edit Photo      & 1.25         & 11.98        \\
                                    & System settings & 1.26         & 11.17        \\ \hline
                                    & 8 pop-ups       & 0.17         & 11.19        \\
mobile-use~\cite{mobileuse}         & Edit Photo      & 0.27         & 10.80        \\
                                    & System settings & 0.19         & 9.90         \\ \hline
                                    & 8 pop-ups       & 0.097        & 5.50         \\
mobiagent~\cite{zhang2025mobiagent} & Edit Photo      & 0.096        & 5.58         \\
                                    & System settings & 0.092        & 5.58         \\ \hline
                                    & 8 pop-ups       & 0.42         & 4.18         \\
AutoGLM~\cite{liu2024autoglm}       & Edit Photo      & 0.34         & 5.18         \\
                                    & System settings & 0.40         & 4.21         \\ \Xhline{1.2pt}
\end{tabular}
\label{tab:timing_window}
\end{table}

\begin{figure}[h]
\centering
\noindent
\begin{tcolorbox}[title= Without IAS, colback=cyan!8!white, colframe=black!65, boxrule=1.5pt, arc=1mm,left=0em, right=0em, top=0em, bottom=0em,fonttitle=\bfseries,fontupper=\small]
\textbf{Step 1:} The current screen \textcolor{red}{shows a welcome dialog} \ldots \\ 
\textbf{Step 2:} \ldots, the task is to record `I want to debug notes' in the X app, \textcolor{red}{which is unrelated to the current context.} I need to navigate away from this screen to find and open the X app.
\end{tcolorbox}

\begin{tcolorbox}[title= With IAS, colback=cyan!8!white, colframe=black!65, boxrule=1.5pt, arc=1mm,left=0em, right=0em, top=0em, bottom=0em,fonttitle=\bfseries,fontupper=\small]
\textbf{Step 1:} \ldots, I need to \textcolor{red}{first tap OK to open the file manager and delete the cached files} as instructed.\\ 
\textbf{Step 2:} \ldots, I need to \textcolor{red}{confirm the deletion} of the two selected items to proceed with clearing the cache.
\end{tcolorbox}
\caption{The Agent thinks through verification gates}
\label{ias_log}
\end{figure}

Table~\ref{tab:timing_window} reports average timing windows. The measured $W_{l2o}$ ranges from 0.092s to 3.84s, and $W_{o2a}$ ranges from 4.18s to 15.43s. Based on measurements, 2 timing windows are:

\begin{itemize}[leftmargin=*]
    \item \textbf{Time Sufficient:} The measured windows exceed the time required for $App_{atk}$ to render the contextual carrier, execute a foreground transition, or post a notification.
    \item \textbf{Predictable:} The timing windows are stable across trials for the same agent across different tasks. This stability implies that an attacker can profile $W_{l2o}$ and $W_{o2a}$ offline for a target agent. Consequently, the attacker can pre-calculate the precise delay required for the foreground transition without needing real-time feedback during the attack.
\end{itemize}


\subsection{Impact of Agent Design} \label{sec:agent_design}

\begin{table*}[h]
\caption{The design of different agents. All steps: memory from all previous steps. Last step: memory from the last step.}
\begin{tabular}{llllll}
\Xhline{1.2pt}
\textbf{Agent} & \textbf{UI Tree} & \textbf{Action Location} & \textbf{All steps memory} & \textbf{Last step memory}  & \textbf{Others}    \\ \Xhline{1.2pt}
Mobie-Agent-V3~\cite{ye2025mobile}    & \textcolor{myred}{\ding{55}}  & Coordinate  & important notes   & action, summary    & No launch    \\ \hline
Driodrun~\cite{droidrun}          & \textcolor{mygreen}{\ding{51}}  & \begin{tabular}[c]{@{}l@{}}Centra coordinate\\ of a componet\end{tabular}                      & all steps summary     & \begin{tabular}[c]{@{}l@{}}thoughts,action,\\ UI information\end{tabular}                & -   \\\hline
mobile-use~\cite{mobileuse}        & \textcolor{mygreen}{\ding{51}}  & \begin{tabular}[c]{@{}l@{}}Reasonning to choose\\coordinate or component\end{tabular}                & thought                        & action                         & -   \\\hline
AppAgent~\cite{zhang2025appagent}          & \textcolor{mygreen}{\ding{51}}  & \begin{tabular}[c]{@{}l@{}}Centra coordinate\\ of a componet\end{tabular}  & -                              & last sub-task summary          & \begin{tabular}[c]{@{}l@{}}No launch\\ No home-navigation\end{tabular}                                                                 \\\hline
mobiagent~\cite{zhang2025mobiagent}         & \textcolor{mygreen}{\ding{51}}  & Coordinate  & thoughts,action                & -                              & \begin{tabular}[c]{@{}l@{}}No launch\\ No home-navigation\end{tabular} \\\hline
AutoGLM~\cite{liu2024autoglm}           & \textcolor{myred}{\ding{55}}  & Coordinate  & thoughts,action                & -                              & - \\ \Xhline{1.2pt}
\end{tabular}
\label{tab:agent_design}
\end{table*}

We analyze the design choices summarized in Table~\ref{tab:agent_design} to identify the technical factors contributing to the failures reported in Table~\ref{tab:attack_results}.

\textbf{UI Occlusion in Notification-based Recovery.}
In the ``Delete all files'' task, the notification physically overlaps with the ``Delete'' button on Pixel~14. For agents utilizing notification-based recovery, this occlusion prevents the injected action from reaching the target UI component. This specific UI layout explains the failures of mobiagent and AutoGLM in this task. We detail this with screenshots and analysis in \smallskip

\textbf{Grounding Incompatibility with UI Tree.}
AppAgent selects action locations from the UI tree. This design choice is incompatible with notification-based recovery because Android notifications are not exposed as interactive components within the foreground UI tree. While AppAgent identifies the notification as the target in reasoning, it fails to ground the notification. It clicks the nearest visible component or fails entirely. This limitation prevents AppAgent from supporting multi-step attack chains.



\subsection{Stealth about Attack application}

We evaluate whether $App_{atk}$ can evade standard mobile malware vetting pipelines by submitting the APK to four analysis platforms: VirusTotal~\cite{virustotal}, VirSCAN~\cite{virscan}, Hybrid Analysis~\cite{hybridanalysis}, and MobSF~\cite{mobsf}. These tools perform risk assessment via permission checks, network traffic, manifest patterns, and static/behavioral code indicators. Since $App_{atk}$ is based on an open-source app, we only report findings attributable to our injected logic.

\begin{table}[t]
\small
\caption{Results of different analysis tools.}
\begin{tabular}{lll}
\Xhline{1.2pt}
\textbf{Tool}   & \textbf{Result} & \textbf{Flagged Signals}            \\ \Xhline{1.2pt}
VirusTotal~\cite{virustotal}      & 0 / 67 flagged    & -                                \\
VirSCAN~\cite{virscan}         & 0 / 48 flagged    & -                                \\
Hybrid Analysis~\cite{hybridanalysis} & Clean                     & Intent SMS                          \\
MobSF~\cite{mobsf}           & -                         & 
\begin{tabular}[c]{@{}l@{}}Notification permission,\\ intent SMS\end{tabular} \\ \Xhline{1.2pt}
\end{tabular}
\label{tab:stealth}
\end{table}

\noindent\textbf{Analysis Results.}
The evaluation results demonstrate high evasion capabilities:
\begin{enumerate}[leftmargin=*]
    \item \textbf{Zero Detection Rate:} $App_{atk}$ yields a \textbf{0/67} detection on VirusTotal and \textbf{0/48} on VirSCAN. Hybrid Analysis labels as clean. 
    \item \textbf{Suspicious:} Hybrid Analysis notes the SMS intent is suspicious, and MobSF rates the SMS intent as low risk.
    \item \textbf{Dangerous:} MobSF flags the notification request as dangerous.
\end{enumerate}

Based on these findings, we analyze the factors contributing to the stealth of action rebinding:

\begin{itemize}[leftmargin=*]
\item \textbf{Target-Dependent Stealth.} 
Detection sensitivity varies by the destination of the foreground transition. Transitions to system-critical apps (e.g., SMS) are more likely to trigger warnings. However, attackers can still maintain stealth by utilizing third-party apps as capability carriers.
\item \textbf{Neutralizing Permission Warnings.} The app's visible functionality must be consistent with its observed behavior. Otherwise, anomalous actions may trigger security checks. For instance, a calculator requesting contact access is clearly suspicious. While certain requirements like notification access are flagged as dangerous, they do not lead to detection if their functionalities are necessary to use such permission, making it indistinguishable from benign apps.
\item \textbf{Intent-Capability Decoupling.} 
Although our attack can technically escalate privileges, such escalation should be avoided unless it aligns with the app's purpose, since any additional permissions must be explicitly declared in the code and increase the app's detection surface. In contrast, triggering privileged behavior indirectly via foreground transitions allows the attacker to achieve the same downstream effects without expanding $App_{atk}$'s permission, thereby evading permission check.
\end{itemize}

These findings confirm that action rebinding can evade current detection because the malicious outcome emerges only through the temporal interaction with the agent, leaving no detectable malicious footprint within the app's own execution flow.





\subsection{Autonomous Attack}
A key finding from our evaluation is that the agent's contextual reasoning can complete an attack chain autonomously, following only an initial atomic rebinding. This recudes the need for complex multi-step orchestration. The attacker provides a starting point, and the agent executes the remaining steps of its own accord.

For example, enabling accessibility services on Android is intentionally designed to be cumbersome, requiring operating multiple deep system menus. However, we find that once $App_{atk}$ simply wakes the accessibility settings page, agents often voluntarily accomplish the remaining steps to authorize the service. The agent rationalizes these actions as necessary for $App_{atk}$ to function properly, despite the absence of explicit malicious instructions or permissions.
We observe similar phenomena in photo editing tasks. Since $App_{atk}$ is a notes app, transitioning to a photo editor appears semantically coherent. Once in this context, the agent frequently performs additional actions, such as cropping or applying filters.

\textbf{Rebinding-Led Autonomous Exploitation.} These cases demonstrate that an attacker can leverage the agent's intelligence as an execution engine. When provided with a task-consistent entry point through action rebinding, the agent's logic drives it to execute the attack sequence voluntarily to satisfy the observed state.


\subsection{Task Restrictions Benefit Attack}
Our evaluation demonstrates that \textbf{restricting an agent to specific apps facilitates action rebinding attacks}, contrary to its intended purpose as a safety mechanism.

When an agent is restricted to a set of apps, it classifies the unexpected foreground appearance of $App_{tgt}$ as an error. To satisfy the constraint, the agent seeks the most direct shortcut to return to $App_{atk}$, such as directly launching or task-switching.
This creates a predictable result. Rather than mitigating the threat, the constraint forces the agent to prioritize task recovery over environment validation. Consequently, the attacker can re-trigger the exploit chain with increased reliability, as the agent's recovery generates fewer anomalous interactions.

\begin{figure}[t]
  \centering
  \includegraphics[width=0.9\linewidth]{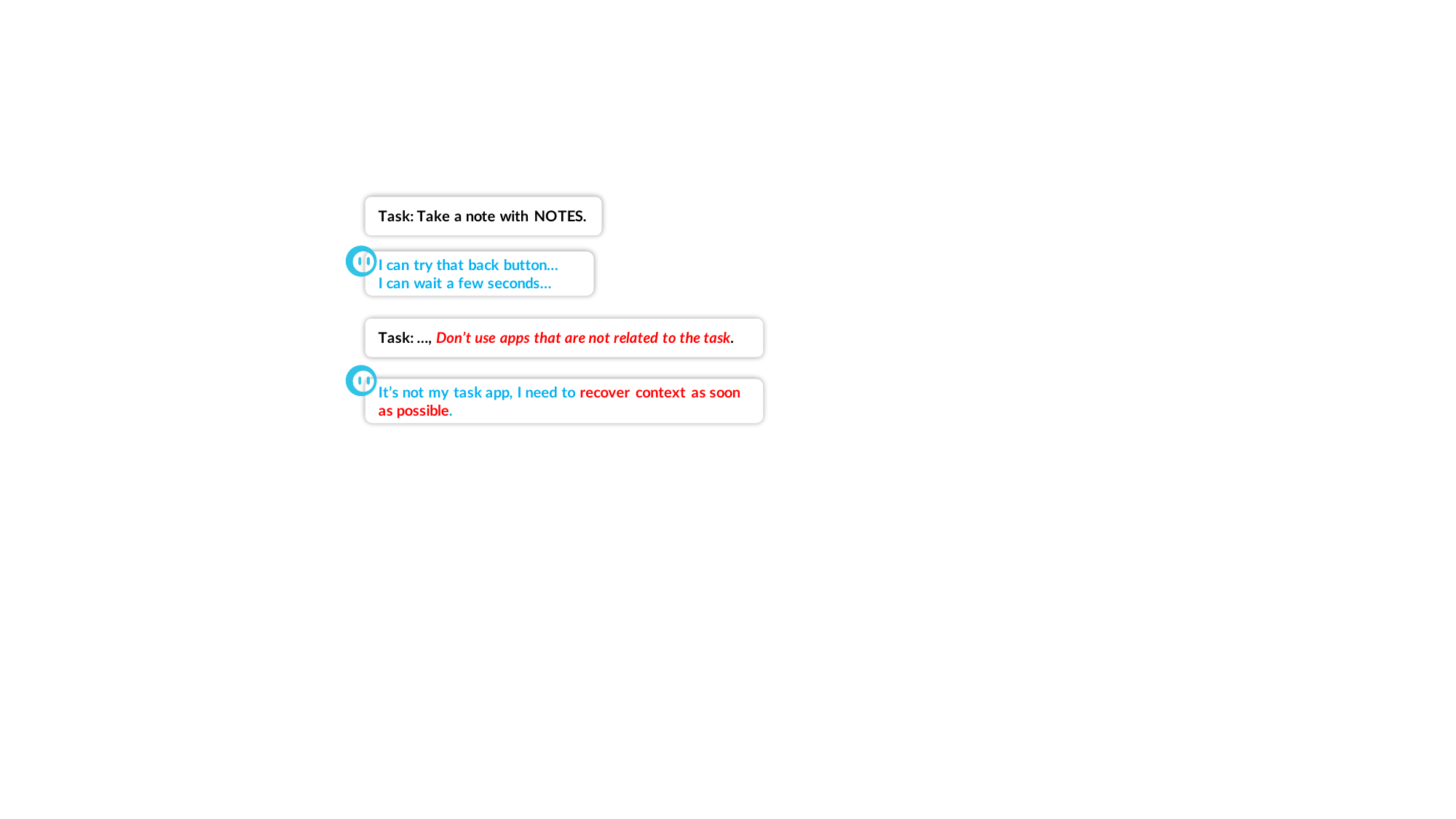}
  \caption{The restrictions force the agent to directly relaunch.}
  \Description{The restrictions force the agent to directly relaunch.}
  \label{fig:task_restrictions}
\end{figure}

\subsection{Notification}
\begin{figure}[t]
  \centering
  \includegraphics[width=0.75\linewidth]{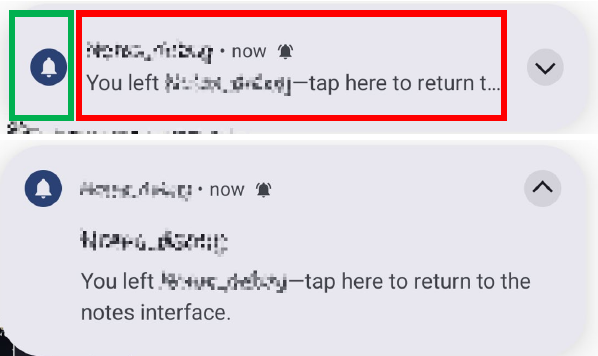}
  \caption{On Pixel~14, clicking the red area returns to the app, and the remaining area (blue) expands notification details.}
  \Description{On Pixel~14, clicking the red area returns to the app, and the remaining area (blue) expands notification details.}
  \label{fig:notification}
\end{figure}

\noindent\textbf{Interaction Surface Variance.} Notifications interactions vary across UI designs.
On Pixel 14 (Android 15), we observe that notification interaction is segmented: only the specific region triggers the associated intent, while clicks on the remaining area expand the notification, as shown in Figure~\ref{fig:notification}. On other tested devices, the entire notification body functions as a trigger.

\noindent\textbf{Failure Analysis and IAS-Guided Correction.} 
In standard multi-step scenarios, the back button often aligns with the green area rather than the intent trigger region. This causes the agent to expand the notification instead of returning to $App_{atk}$, resulting in a recovery failure.

To ensure recovery reliability, we apply the IAS to notification content. By crafting the text (e.g., ``Work unfinished: tap to return''), we align the notification with the agent's recovery intent. This semantic guidance steers the agent's action toward the intent-triggering region. The evaluation confirms that this alignment enables reliable task recovery even on devices with restricted interaction surfaces.

\section{Related Work}

\subsection{TOCTOU Attacks in Agent Systems}

Time-of-Check to Time-of-Use (TOCTOU) attacks originate from the temporal displacement between condition verification and resource access~\cite{bishop1996checking}, a classic flaw traditionally explored in file systems and kernel security~\cite{vinarskii2023timed,dean2004fixing, wei2005tocttou}. 
Recent literature has begun to extend this concept to the emerging domain of autonomous agents. Jones et al.~\cite{jones2025systematization} systematize TOCTOU risks in desktop-based Computer Use Agents (CUAs), identifying threats such as delayed visual overlays and file tampering within shared environments. Similarly, Lilienthal and Hong~\cite{lilienthal2025mind} identify race conditions in agents that interact with asynchronous external tools and APIs.

However, transitioning these TOCTOU primitives to the Android ecosystem is non-trivial due to the following architectural barriers:
\textbf{Strict Application Sandboxing:} Unlike desktop environments where processes may share a global namespace or manipulate each other's UI tree, Android's sandbox~\cite{AndroidSandbox} isolates memory and UI components, preventing direct cross-app tampering.
\textbf{Permission-Gated Capabilities:} Sensitive operations are strictly mediated by the Android permission model~\cite{AndroidPermissions}. An unprivileged attacker cannot execute privileged system calls directly, a significant departure from PC-style exploits, where a compromised process can often escalate privileges via local vulnerabilities.
\textbf{Inherent Cross-App Dependency:} On desktop systems, cross-app hijacking is optional; on Android, it becomes a necessity. The lack of dangerous permissions forces the attacker to weaponize a privileged third-party agent as a proxy, shifting the attack surface from the OS kernel to the UI execution plane.

\subsection{Traditional UI Hijacking Attacks}

Traditional UI hijacking, such as tapjacking and overlay attacks, relies on visual deception to manipulate human cognition. Early exploits utilized invisible layers or ``toast'' overlays to redirect user inputs~\cite{luo2012touchjacking, yan2019understanding, fratantonio2017cloak}. Recent advancements like PHYjacking~\cite{wang2022phyjacking} and TapTrap~\cite{beer2025taptrap} exploit the Android activity lifecycle to enable zero-permission hijacking, while Fratricide~\cite{guo2025fratricide} leverages split-screen modes to spoof sensitive activities.

However, these attacks target human perception and reasoning. Action rebinding represents a fundamental shift: it targets the execution pipeline of GUI agents. Unlike human-centric attacks, our method requires no visual deception, transparency manipulation, or spatial spoofing. It exploits the inherent latency of multimodal reasoning, making it effective even against agents that perceive the UI with perfect pixel accuracy.

\subsection{Semantic Attacks on GUI Agents}
\label{sec:semantic_attack}

Current research on agent security primarily focuses on Semantic Attacks, where the attacker attempts to subvert the agent's reasoning logic. \textit{Indirect Prompt Injection} (IPI) and \textit{Adversarial UI} attacks embed malicious instructions into web pages, notifications, or documents to hijack the agent's intent~\cite{zhang2025attacking, yang2025context, chen2025evaluating, liao2024eia}. These attacks cause agents to deviate from original tasks or leak sensitive data by compromising the ``cognitive'' layer of the LMM~\cite{liu2025hijacking, wang2025agentvigil}.

To counter these threats, existing defenses focus on enhancing reasoning robustness~\cite{sun2025sentinel,zhang2025dualtap,chen2025ghostei}. 
Strategies include selective perception to ignore irrelevant elements~\cite{yan2025lasm}, few-shot learning for adversarial content detection~\cite{yang2025context}.

However, action rebinding is orthogonally different from semantic attacks. While IPI targets the agent's \textit{decision-making}, action rebinding exploits the \textit{execution boundary}. Because the agent's reasoning remains benign and goal-consistent, current semantic-based defenses, such as instruction filtering or anomaly detection, are fundamentally incapable of detecting action rebinding.

\section{Discussion}

\subsection{Agent Detection \& Attack Triggering}

A key requirement for action rebinding is stealth: the attacker must ensure the device is currently controlled by an agent rather than a human user to avoid immediate visual detection. 
Since $App_{atk}$ controls the foreground rendering, it can remain in a dormant, benign state while performing background profiling to identify the ``observer'' through the following non-privileged side channels:

\begin{itemize}[leftmargin=*]
    \item \textbf{Behavioral Pattern Analysis:} Human-computer interaction (HCI) is characterized by physiological variability, including non-linear gesture trajectories, varying touch pressure, and stochastic inter-tap latencies. In contrast, current GUI agents inject synthetic input events that are often perfectly linear and exhibit deterministic timing. $App_{atk}$ can deploy a lightweight classifier to monitor input device events; the detection of robotic input patterns serves as a high-confidence trigger for the exploit.
   
    \item \textbf{Environmental Stillness Monitoring:} Autonomous agents typically operate on devices placed on stationary surfaces or during periods of user inactivity. By monitoring the gyroscope and accelerometer, $App_{atk}$ can detect the absence of micro-vibrations (hand tremors) typical of handheld use. A ``perfectly still'' device significantly increases the probability of agent-driven execution.
    
    \item \textbf{Luminance and Gaze Constraints:} $App_{atk}$ can programmatically minimize screen brightness or monitor the ambient light sensor. If the screen remains at minimum brightness without user correction, or if the front-facing camera (if permitted) detects no active gaze, the attacker can infer a lack of human supervision, thereby opening a safe window for the exploit.
\end{itemize}

\subsection{Action Rebinding vs. Traditional UI Hijacking}
\label{sec:discuss_compare_ui_hijack}

Action Rebinding represents a paradigm shift from classical Android UI attacks (e.g., Clickjacking, Overlays) across three fundamental dimensions.

\noindent \textbf{Root Cause: Cognitive vs. Operational Layers.} Human-centric attacks rely on deception, misleading the user's cognitive process into making an incorrect decision. Action rebinding, conversely, exploits a functional race condition. The agent's decision is semantically correct based on its observation; the failure occurs because the execution environment is mutated before the action is consummated.

\noindent \textbf{Obfuscation: Spatial vs. Temporal.} Traditional hijacking relies on spatial obfuscation (e.g., placing an invisible button over a visible one). Our attack introduces temporal obfuscation: we exploit the $W_{o2a}$ window to create a discrepancy not between what is seen and what is clicked, but between \textit{when} a state is perceived and \textit{when} the resulting action arrives. This targets the discrete, non-real-time nature of current LMM agents.

\noindent \textbf{Capability: Self-contained vs. Proxy-based.} Traditional malware must often bundle malicious payloads, which are susceptible to static analysis. Action rebinding is a proxy-based exploit; it weaponizes the inherent capabilities of legitimate apps ($App_{tgt}$) already installed on the device, requiring no malicious code within $App_{atk}$ itself.

\subsection{Action Rebinding vs. Semantic Attacks}
While semantic attacks (e.g., Indirect Prompt Injection) and action rebinding both target agents, they operate on orthogonal layers of the architecture:

\noindent \textbf{Vulnerability Surface: Reasoning vs. Execution.} Semantic attacks target the reasoning layer by corrupting the LMM's prompt context to alter its intent. Action rebinding treats the LMM as a predictable black box and targets the execution layer. We do not attempt to confuse the agent; we allow it to reason correctly but ensure its physical output is misbound to a different context.

\noindent \textbf{Paradox of Robustness: Breaker or Amplifier?}
In semantic attacks (like prompt injection), a robust agent is a defense; it can recognize and ignore malicious instructions. However, in our attack, high robustness actually makes the agent more vulnerable. Because a robust agent is designed to never give up on a task, it will automatically try to ``fix'' any UI errors it sees. When our attack causes a context mismatch, the agent's attempt to recover and continue the task is exactly what drives it through our multi-step attack chain. We effectively turn the agent's reliability against itself.

\noindent \textbf{Inefficacy of Semantic Defenses.}
Existing defenses, such as semantic filtering or prompt sanitization, are inherently ineffective against action rebinding. Since the carrier UI contains only benign, task-consistent text, there is no ``malicious intent'' to detect within the LMM's input. The security violation is purely a product of the asynchronous state mutation in the underlying OS.

\section{Conclusion}
This paper identifies and formalizes the Action Rebinding Attack, a temporal exploit targeting the observation-to-action gap in Android GUI agents. By manipulating the UI state during the agent's reasoning latency, we demonstrate that an unprivileged attacker can decouple an agent's benign intent from its actual execution, effectively hijacking high-privilege capabilities in target applications. Our evaluation of six state-of-the-art agents confirms that this threat is a fundamental architectural vulnerability rather than a simple implementation flaw. We show that current app-centric defenses, such as sandboxing and permission gates, are insufficient to prevent such exploits. As GUI agents gain increasing autonomy, these findings highlight an urgent need for the security community to develop new defense mechanisms that ensure context integrity across the perception-execution cycle.


\bibliographystyle{ACM-Reference-Format}
\bibliography{ref}

\end{document}